\documentclass[table]{ccjnl}
\usepackage{makecell}
\usepackage{lipsum,amsmath}
\usepackage{cuted}
\usepackage{bm}
\usepackage{stfloats}
\graphicspath{{figures/}}
\usepackage{color}

\title{A Realistic 3D Non-Stationary Channel Model for UAV-to-Vehicle Communications Incorporating Fuselage Posture}
\author{Boyu Hua\inst{1}, Tongtong Zhou\inst{1}, Qiuming Zhu\inst{1,*},Kai Mao\inst{1}, Junwei Bao\inst{2}, Weizhi Zhong\inst{3}, Naeem Ahmed\inst{1}\corinfo{zhuqiuming@nuaa.edu.cn}}
\receiveddate{19-Apr-2021}
\reviseddate{16-June-2021, accepted at 28-Oct-2021}
\Editor{Honggang Zhang}

\address[1]{College of Electronic and Information Engineering, Nanjing University of Aeronautics and Astronautics, Nanjing 210016, China}
\address[2]{College of Science, Nanjing University of Aeronautics and Astronautics, Nanjing 210016, China}
\address[3]{College of Astronautics, Nanjing University of Aeronautics and Astronautics, Nanjing 210016, China}

\begin{document}
\maketitle
\begin{abstract}
Considering the unmanned aerial vehicle (UAV) three-dimensional (3D) posture, a novel 3D non-stationary geometry-based stochastic model (GBSM) is proposed for multiple-input multiple-output (MIMO) UAV-to-vehicle (U2V) channels. It consists of a line-of-sight (LoS) and non-line-of-sight (NLoS) components. The factor of fuselage posture is considered by introducing a time-variant 3D posture matrix. Some important statistical properties, i.e. the temporal autocorrelation function (ACF) and spatial cross correlation function (CCF), are derived and investigated. Simulation results show that the fuselage posture has significant impact on the U2V channel characteristic and aggravate the non-stationarity. The agreements between analytical, simulated, and measured results verify the correctness of proposed model and derivations. Moreover, it is demonstrated that the proposed model is also compatible to the existing GBSM without considering fuselage posture.
\end{abstract}
\section{INTRODUCTION}
\label{Introduction}
Owing to the versatility and high mobility, the unmanned aerial vehicles (UAVs) have been considered to be a promising new paradigm to facilitate scenarios including communication relay, hazardous exploration, traffic control, and disaster management \cite{ChengX19JCIN,FeiZS19IoTJ,ZhongWZ19ChCom}. Compared with the conventional vehicle-to-vehicle communication, the UAV-to-vehicle (U2V) communication has some unique channel characteristics due to the UAV flight, e.g., three-dimensional (3D) scattering space, 3D arbitrary trajectory, 3D antenna arrangement, and 3D fuselage posture\cite{ZhuQM19WCL}. These new features would cause different non-stationarity from traditional communication channel, which conventional channel models cannot describe appropriately. To better design and evaluate the future U2V communication systems, it is essential to establish a realistic and reliable U2V channel model \cite{Ullah20TCCN,GuanK19MAP,ZhuQM20IWCMC}.

In past decades, there were quite a lot of researches on the UAV channel modeling. These works can be classified into two categories, i.e., the small-scale fading model and large-scale fading model \cite{Khawaja19CST}. The former studied the fast time-variant parameters like Doppler shift and channel fading, while the latter focused the slow time-variant parameters such as the path loss and shadowing \cite{FengW20WC,FengW19IoTJ,AlHourani18WCL}. Due to the fast time-variant and non-stationary characteristics of U2V communication scenarios, the small-scale fading becomes more severe. Among the existing small-scale fading channel models, the geometry based stochastic model (GBSM) has been widely accepted \cite{WangWM20IJEC,ZhaoXW19ChCom,ZhuQM20EuCAP}, which has moderate complexity and accuracy compared with deterministic and other stochastic models. For example, a basic 3D stationary GBSM was proposed in \cite{Gulfam16AS} to study the UAV channels. The authors in \cite{JiangLG17CL} considered the channel non-stationarity by adding the dynamic evolution of scatterers into the model. These models assumed the terminals were fixed, which limited the versatility of such kind of models. Different from these basic UAV channel models, U2V channel models focus on the movements of both UAV and ground terminal, and has become a significant research issue.

Some of existing U2V channel models assumed that the UAV or vehicle moves at uniform speed in a straight line \cite{ZhangZC18CL,GuanK19VTC, WangCX17TWC}. However, in a practical environment, the vehicle may experience velocity change and arbitrary trajectory. The non-stationary U2V channel models in \cite{Borhani17TVT,ZhuQM20Access} considered both the UAV and vehicle moved with fixed velocities, which was not consist with the practical facts. Some modified non-stationary channel model was proposed by taking the 3D speed variation of UAV into considerition \cite{WangCX20IoTJ,ChengX19IETCom1,ChengX19IETCom2}, but the movement of ground terminal is fixed. By setting the fixed trajectory, authors in \cite{GuanK19AWPL,GuanK19ISAP,HeDP18EuCAP} proposed more realistic UAV channel models and studied the related simulations. Moreover, the authors in \cite{ZhuQM18ChCom,ZhuQM19MAP} proposed a U2V multiple-input multiple-output (MIMO) channel model and the transceivers moved along 3D trajectories. However, the time-variant fuselage posture rotation, which is inevitable in the realistic scenario, was not involved.

Note that the aforementioned U2V channel models considered the linear motion, curved motion or 3D arbitrary motion of UAVs. However, the posture rotation (including roll, pitch, and yaw) of the UAV was not been considered. Recently, the impact of drone pitch was investigated in \cite{AiB20TVT}, some corresponding statistical properties were derived. However, to the best of the authors’ knowledge, the thoroughly study of 3D posture and the realistic U2V channel model incorporating fuselage posture is still missing. This paper aims to fill this research gap. The main contributions and innovations of this paper are summarized as follows:

  1) A realistic non-stationary U2V MIMO channel model for 3D scattering space, 3D arbitrary trajectory, 3D antenna arrangement, and 3D fuselage posture is proposed. It pays the special consideration of UAV's 3D posture rotation, i.e. pitch, yaw, and roll. Thus, the non-stationarity caused by the fuselage posture is taken into account. In order to support 3D trajectory and rotation of UAV, this paper introduces the rotation matrix to model parameters that related with the time-variant velocity and posture.

  2) The expressions of some key statistical properties, i.e., the temporal autocorrelation function (ACF) and spatial cross correlation function (CCF) are obtained and verified by the analytical and simulation results. Besides, different influence on channel statistical properties caused by the UAV posture are simulated and discussed. The observations and conclusions can be used as a reference for the system design and performance analysis of U2V MIMO communication systems.

  3) The generality of proposed non-stationary U2V channel model is validated by comparing the statistical properties with measurement results in different scenarios. The simulated results have a good consistency with the measured results. Therefore, the presented realistic model can be adopted to diverse UAV communication scenarios by adjusting model parameters.

The remainder of this paper is organized as follows. In Section~\ref{sec:U2V Channel Model}, a new 3D non-stationary channel model for U2V communications incorporating fuselage posture is presented. Section~\ref{sec:Statistical Property Analysis of Proposed Model} studies the spatial-temporal correlation and the rotation matrix with effective phase, along with two typical statistical properties of proposed model. Section~\ref{sec:Numerical Results and Discussions} compares and discusses the analytical and simulation results. Finally, conclusions are drawn in Section~\ref{sec:Conclusions}.

\section{U2V CHANNEL MODEL INCORPORATING FUSELAGE POSTURE}
\label{sec:U2V Channel Model}
\subsection{Channel Models Comparison}
To illustrate the contribution of this paper, the proposed model, the model in \cite{ZhuQM18ChCom,ZhuQM19MAP}, and the standardized models, i.e., ITU-R \cite{A}, METIS \cite{B},mmMAGIC \cite{D}, IEEE 802.11 ad \cite{C}, 3GPP \cite{3GPP2020}, and QuaDRiGa \cite{F}, are investigated and compared in Table 1. The supported frequency band and scenario are listed in numerical form. The features like MIMO, 3D propagation, dual mobility, non-stationarity, and fuselage posture are displayed with a tick or cross. Note that most of standardized channel models cannot support both the non-stationarity and fuselage posture, the proposed model aims to fill this gap.

\begin{table*}[htb]
\caption{Comparison with different channel models.}
\setlength{\tabcolsep}{3pt}
\begin{tabular}{|p{100pt}|p{70pt}|p{80pt}|p{40pt}|p{40pt}|p{40pt}|p{40pt}|p{40pt}|}

\hline
\makecell[c]{Standards}& \makecell[c]{ Frequency}&\makecell[c]{Scenario}&\makecell[c]{MIMO}&3D pro-

pagation&Dual 

mobility&Non-stationary&Fuselage posture\\
\hline
\makecell[c]{ITU-R M \cite{A}}&\makecell[c]{Up to 6 GHz}&\makecell[c]{Dense urban}& \makecell[c]{$\surd$} & \makecell[c]{$\surd$}	&\makecell[c]{$\times$}	&\makecell[c]{$\times$}	&\makecell[c]{$\times$}\\
\hline
\makecell[c]{METIS \cite{B}}&	\makecell[c]{2-60 GHz}	&\makecell[c]{Outdoor/indoor}&	\makecell[c]{$\surd$}&	\makecell[c]{$\surd$	}&\makecell[c]{$\times$}	&\makecell[c]{$\times$}&	\makecell[c]{$\times$}\\
\hline
\makecell[c]{mmMAGIC \cite{D}}& \makecell[c]{6-100 GHz}&	\makecell[c]{Outdoor/indoor}&\makecell[c]{	$\surd$	}&\makecell[c]{$\surd$}	&\makecell[c]{$\times$}&	\makecell[c]{$\times$}&	\makecell[c]{$\times$}\\
\hline
\makecell[c]{IEEE 802.11 ad \cite{C}}&\makecell[c]{Up to 60 GHz}&\makecell[c]{	Indoor}	&\makecell[c]{$\surd$}&	\makecell[c]{$\surd$}&	\makecell[c]{$\surd$}&	\makecell[c]{$\times$}&	\makecell[c]{$\times$}\\
\hline
\makecell[c]{3GPP TR 38.901 \cite{3GPP2020}}&\makecell[c]{0.5-100 GHz}&	\makecell[c]{Outdoor/indoor}&	\makecell[c]{$\surd$	}&\makecell[c]{$\surd$}	&\makecell[c]{$\surd$}&	\makecell[c]{$\times$}&\makecell[c]{	$\times$}\\
\hline
\makecell[c]{QuaDRiGa \cite{F}}&\makecell[c]{Up to 100 GHz}&	\makecell[c]{Indoor/satellite}	&\makecell[c]{$\surd$}	&\makecell[c]{$\surd$}	&\makecell[c]{$\surd$	}&\makecell[c]{$\surd$	}&\makecell[c]{$\times$}\\
\hline
\makecell[c]{Model in \cite{ZhuQM18ChCom,ZhuQM19MAP}}&	\makecell[c]{0.5-100 GHz}&	\makecell[c]{Outdoor	}&\makecell[c]{$\surd$}&	\makecell[c]{$\surd$}&	\makecell[c]{$\surd$}	&\makecell[c]{$\surd$}	&\makecell[c]{$\times$}\\
\hline
\makecell[c]{The proposed model}&	\makecell[c]{0.5-100 GHz}&	\makecell[c]{Outdoor	}&\makecell[c]{$\surd$}	&\makecell[c]{$\surd$}&	\makecell[c]{$\surd$}	&\makecell[c]{$\surd$}&\makecell[c]{	$\surd$}\\
\hline
\end{tabular}
\label{table1}
\end{table*}

\subsection{Channel Impulse Response of Porposed Model}
\begin{figure}[htbp]
  \centering{\includegraphics[width=0.5\textwidth]{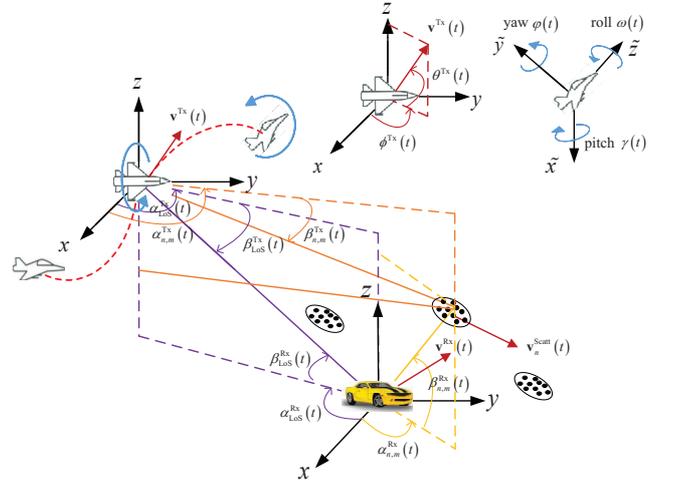}}
  \caption{3D non-stationary GBSM for U2V channels.} 
  \label{fig1} 
\end{figure}

The proposed non-stationary MIMO GBSM for U2V communication systems is as shown in Fig. \ref{fig1}. The mobile transmitter (Tx), i.e. the UAV, is equipped with $P$ antennas, while the mobile receiver (Rx), i.e. the vehicle on the ground, is equipped with $Q$ antennas. The proposed model consists of a line-of-sight (LoS) component and non-line-of-sight (NLoS) components that are equivalent to the single bounce components in this case. It should be noted that the Tx and Rx belong to two different coordinate systems, which are respectively marked as $\widetilde{x}\text{ - }\widetilde{y}\text{ - }\widetilde{z}$ coordinate and $x\text{ - }y\text{ - }z$ coordinate. The two original positions are the center of Tx and Rx, respectively. To describe the posture of fuselage, three posture angles, i.e. roll, pitch, and yaw, are introduced in the coordinate. The detailed definitions of parameters in Fig. \ref{fig1} are listed in Table \ref{table2}.

\begin{table}[htb]
\caption{Parameter definitions in the proposed model}
\setlength{\tabcolsep}{3pt}
\begin{tabular}{|p{80pt}|p{150pt}|}
\hline
${{\mathbf{v}}^{\text{Tx}}}\left( t \right)$,${{\mathbf{v}}^{\text{Rx}}}\left( t \right)$, $\mathbf{v}_{n}^{\text{Scatt}}\left( t \right)$&
3D velocity vectors of the Tx, Rx, and scatterers\\
\hline
& \\[-10pt]  
$\alpha _{\text{LoS}}^{\text{Tx}}\left( t \right)$, $\alpha _{\text{LoS}}^{\text{Rx}}\left( t \right)$&
Azimuth angle of departure and arrival for Tx and Rx in the LoS path\\
\hline
& \\[-10pt]
$\beta _{\text{LoS}}^{\text{Tx}}\left( t \right)$, $\beta _{\text{LoS}}^{\text{Rx}}\left( t \right)$&
Elevation angle of departure and arrival for Tx and Rx in the LoS path\\
\hline
& \\[-10pt]
$\alpha _{n,m}^{\text{Tx}}\left( t \right)$, $\alpha _{n,m}^{\text{Rx}}\left( t \right)$&
Azimuth angle of departure and arrival in the NLoS paths\\
\hline
& \\[-10pt]
$\beta _{n,m}^{\text{Tx}}\left( t \right)$, $\beta _{n,m}^{\text{Rx}}\left( t \right)$&
Elevation angle of departure and arrival for Tx and Rx in the NLoS paths\\
\hline
& \\[-10pt]
${{\phi }^{\text{Tx}}}\left( t \right)$, ${{\phi }^{\text{Rx}}}\left( t \right)$&
Azimuth angle of the velocity vector of Tx and Rx, respectively\\
\hline
& \\[-10pt]
${{\theta }^{\text{Tx}}}\left( t \right)$, ${{\theta }^{\text{Rx}}}\left( t \right)$&
Elevation angle of the velocity vector of Tx and Rx, respectively\\
\hline
& \\[-10pt]
$\omega \left( t \right)$, $\gamma \left( t \right)$, $\varphi \left( t \right)$&
Roll, pitch, and yaw angle of the UAV posture, respectively\\
\hline
\end{tabular}
\label{table2}
\end{table}

The U2V MIMO channel between the UAV installed $P$ antenna elements and the vehicle installed $Q$ antenna elements can be defined as a complex matrix \cite{ZhuQM18TC}, i.e.,
\begin{equation}\text{H}\left( t,\tau  \right)\!=\!{{\left[ \begin{matrix}
   {{h}_{1,1}}\left( t,\tau  \right)\! & \!{{h}_{1,2}}\left( t,\tau  \right)\! & \!\cdots \! & \!{{h}_{1,P}}\left( t,\tau  \right)  \\
   {{h}_{2,1}}\left( t,\tau  \right)\! &\! {{h}_{2,2}}\left( t,\tau  \right)\! &\! \cdots\!  &\! {{h}_{2,P}}\left( t,\tau  \right)  \\
   \vdots \! & \!\vdots \! & \!\ddots  \!& \!\vdots   \\
   {{h}_{Q,1}}\left( t,\tau  \right)\! &\! {{h}_{Q,2}}\left( t,\tau  \right) \!& \!\cdots \! & \!{{h}_{Q,P}}\left( t,\tau  \right)  \\
\end{matrix} \right]}_{Q\!\times\! P}}\label{eq1}\end{equation}
where ${{h}_{qp}}\left( t,\tau  \right)$ denotes the complex channel impulse response (CIR) between the $p$-th transmitting antenna element and the $q$-th receiving antenna element. Moreover, the single CIR can be modeled as a superposition of a LoS path component and several NLoS path components, i.e.,
\begin{equation}\begin{aligned}
 & {{h}_{qp}}\left( t,\tau  \right)\text{=}\sqrt{\frac{K}{K+1}}{{B}^{\text{LoS}}}\left( t \right)h_{qp}^{\text{LoS}}\left( t \right)\delta \left( \tau -{{\tau }^{\text{LoS}}}\left( t \right) \right) \\ 
 & +\sqrt{\frac{1}{K+1}}\sum\limits_{n=1}^{N\left( t \right)}{B_{n}^{\text{NLoS}}\left( t \right)h_{qp,n}^{\text{NLoS}}\left( t \right)\delta \left( \tau -\tau _{n}^{\text{NLoS}}\left( t \right) \right)}  
\end{aligned}\label{eq2}\end{equation}
where $K$ denotes the Ricean factor in LoS path, ${{B}^{\text{LoS}}}\left( t \right)$ and $B_{n}^{\text{NLoS}}\left( t \right)$ are variables reflecting the birth-death process of clusters in NLoS paths, ${{\tau }^{\text{LoS}}}\left( t \right)$ and $\tau _{n}^{\text{NLoS}}\left( t \right)$ denote the path delays, $N\left( t \right)$ denotes the number of valid scattering paths, the complex channel coefficient $h_{qp}^{\text{LoS}}\left( t \right)$ and $h_{qp,n}^{\text{NLoS}}\left( t \right)$ can be expressed as (\ref{eq3}) and (\ref{eq4}),
\begin{figure*}[htb]
\small
\begin{equation}
h_{qp}^{\text{LoS}}\left( t \right)={{\left[ \begin{matrix}
   F_{p,\text{V}}^{\text{Tx}}\left( \alpha _{\text{LoS}}^{\text{Tx}}\left( t \right),\beta _{\text{LoS}}^{\text{Tx}}\left( t \right) \right)  \\
   F_{p,\text{H}}^{\text{Tx}}\left( \alpha _{\text{LoS}}^{\text{Tx}}\left( t \right),\beta _{\text{LoS}}^{\text{Tx}}\left( t \right) \right)  \\
\end{matrix} \right]}^{T}}\left[ \begin{matrix}
   1 & 0  \\
   0 & -1  \\
\end{matrix} \right]\left[ \begin{matrix}
   F_{q,\text{V}}^{\text{Rx}}\left( \alpha _{\text{LoS}}^{\text{Rx}}\left( t \right),\beta _{\text{LoS}}^{\text{Rx}}\left( t \right) \right)  \\
   F_{q,\text{H}}^{\text{Rx}}\left( \alpha _{\text{LoS}}^{\text{Rx}}\left( t \right),\beta _{\text{LoS}}^{\text{Rx}}\left( t \right) \right)  \\
\end{matrix} \right]{{e}^{\text{j}\left( \Phi _{\text{I}}^{\text{LoS}}\left( t \right)\text{+}\Phi _{\text{D}}^{\text{LoS}}\left( t \right)+\Phi _{{{\text{A}}_{qp}}}^{\text{LoS}}\left( t \right) \right)}}
\label{eq3}\end{equation}
\end{figure*}
\begin{figure*}[htb]
\footnotesize
\begin{equation}
h_{qp,n}^{\text{NLoS}}\!\left( t \right)\text{=}\!\underset{M\to \infty }{\mathop{\lim }}\,\!\!\!\sqrt{\!\frac{1}{M}}\!\!\!\sum\limits_{m=1}^{M}{{{\!\!\left[ \begin{matrix}
   F_{p,\!\text{V}}^{\text{Tx}}\!\left( \alpha _{\text{n,m}}^{\text{Tx}}\!\left( t \right)\!,\beta _{\text{n,m}}^{\text{Tx}}\!\left( t \right) \right)  \\
   F_{p,\!\text{H}}^{\text{Tx}}\!\left( \alpha _{\text{n,m}}^{\text{Tx}}\!\left( t \right)\!,\beta _{\text{n,m}}^{\text{Tx}}\!\left( t \right) \right)  \\
\end{matrix} \right]}^{T}}\!\!\left[ \begin{matrix}
   {{e}^{\text{j}\Phi _{n,m}^{\text{VV}}}} \!\!&\!\!\!\! \sqrt{\!{{\kappa }_{n\!,\!m}}^{\!\text{-}1}}{{e}^{\text{j}\Phi _{n\!,\!m}^{\text{VH}}}}  \\
   \sqrt{{{\kappa }_{n\!,\!m}}^{\!\text{-}1}}{{e}^{\text{j}\Phi _{n\!,\!m}^{\text{HV}}}}\!\!\! &\!\!\!\!\! {{e}^{\text{j}\Phi _{n,m}^{\text{HH}}}}  \\
\end{matrix} \right]\!\!\!\left[ \begin{matrix}
   F_{q,\!\text{V}}^{\text{Rx}}\!\left( \alpha _{\text{n,m}}^{\text{Rx}}\!\left( t \right)\!,\beta _{\text{n,m}}^{\text{Rx}}\!\left( t \right) \right)  \\
   F_{q,\!\text{H}}^{\text{Rx}}\!\left( \alpha _{\text{n,m}}^{\text{Rx}}\!\left( t \right)\!,\beta _{\text{n,m}}^{\text{Rx}}\!\left( t \right) \right)  \\
\end{matrix} \right]{{e}^{\text{j}\left( \!\Phi _{{{\text{I}}_{n\!,\!m}}}^{\text{NLoS}}\!\left(\! t \!\right)\text{+}\Phi _{{{\text{D}}_{n\!,\!m}}}^{\text{NLoS}}\!\left(\! t \!\right)\text{+}\Phi _{{{\text{A}}_{q\!p\!,\!n\!,\!m}}}^{\text{NLoS}}\!\left( \!t\! \right) \!\right)}}}
\label{eq4}\end{equation}
\end{figure*}
where $M$ is the number of sub-paths in a path, $F_{p,\text{V}}^{\text{Tx}}$, $F_{p,\text{H}}^{\text{Tx}}$, $F_{q,\text{V}}^{\text{Rx}}$ and $F_{q,\text{H}}^{\text{Rx}}$ denote the vertically and horizontally polarized field components of the transmitter or receiver, respectively. For each $m$-th sub-path in the $n$-th path, $\Phi _{n,m}^{\text{VV}}$, $\Phi _{n,m}^{\text{VH}}$,$\Phi _{n,m}^{\text{HV}}$ and $\Phi _{n,m}^{\text{HH}}$ are random initial phases for four different polarization combinations, which are uniformly distributed within $\left(-\pi,\pi\right)$, and ${{\kappa }_{n,m}}$ is the cross polarization power ratio.
 Besides, $\Phi _{{{\text{I}}_{n,m}}}^{\text{LoS/NLoS}}\left( t \right)$ is the random initial phase which obeys the uniform distribution over $\left( 0,\text{ }2\pi  \right]$, $\Phi _{{{\text{D}}_{n,m}}}^{\text{LoS/NLoS}}\left( t \right)$ is the time-variant Doppler phase caused by Doppler frequency variation, which can be derived as
\begin{equation}
\Phi _{{{\text{D}}_{{}}}}^{\text{LoS}}\left( t \right)\!\text{=}k\!\!\int_{0}^{t}{\!\left(\! {{\mathbf{v}}^{\text{Tx}}}\left( t' \right)\!\cdot\! \mathbf{s}_{\text{LoS}}^{\text{Tx}}\left( t' \right)\text{+}{{\mathbf{v}}^{\text{Rx}}}\left( t' \right)\!\cdot\! \mathbf{s}_{\text{LoS}}^{\text{Rx}}\left( t' \right)\!\right)\!\text{ }}\text{d}t'
\label{eq5}\end{equation}

\begin{equation}\Phi _{{{\text{D}}_{n,m}}}^{\text{NLoS}}\left( t \right)\!=\!k\!\int_{0}^{t}{\left[ \begin{aligned}
  & \left( {{\mathbf{v}}^{\text{Tx}}}\left( t' \right)\!-\!\mathbf{v}_{n}^{\text{Scatt}}\left( t' \right) \right)\!\cdot\! \mathbf{s}_{n,m}^{\text{Tx}}\left( t' \right) \\ 
 & \text{+}\left( {{\mathbf{v}}^{\text{Rx}}}\left( t' \right)\!-\!\mathbf{v}_{n}^{\text{Scatt}}\left( t' \right) \right)\!\cdot\! \mathbf{s}_{n,m}^{\text{Rx}}\left( t' \right) \\ 
\end{aligned} \right]}\text{ d}t'\label{eq6}\end{equation}
where $k=2\pi {{f}_{0}}/{{c}_{0}}$ denotes the wave number, ${{f}_{0}}$ and ${{c}_{0}}$ represent the operating frequency and speed of electromagnetic wave. $\mathbf{s}_{\text{LoS}}^{\text{Tx}}\left( t \right)$, $\mathbf{s}_{\text{LoS}}^{\text{Rx}}\left( t \right)$, $\mathbf{s}_{n,m}^{\text{Tx}}\left( t \right)$, and $\mathbf{s}_{n,m}^{\text{Rx}}\left( t \right)$ are the departure and arrival angle unit vectors of the LoS path and the $m$-th sub-path within the $n$-th NLoS path, respectively. Furthermore, they can be defined by
\begin{equation}
\mathbf{s}_{\text{LoS}/n,m}^{\text{Tx}/\text{Rx}}\!\left( t \right)\!=\!\left[ \begin{matrix}
   \cos \beta _{\text{LoS/}n,m}^{\text{Tx}/\text{Rx}}\left( t \right)\cos \alpha _{\text{LoS/}n,m}^{\text{Tx}/\text{Rx}}\left( t \right)  \\
   \cos \beta _{\text{LoS/}n,m}^{\text{Tx}/\text{Rx}}\left( t \right)\sin \alpha _{\text{LoS/}n,m}^{\text{Tx}/\text{Rx}}\left( t \right)  \\
   \sin \beta _{\text{LoS/}n,m}^{\text{Tx}/\text{Rx}}\left( t \right)  \\
\end{matrix} \right].
\label{eq7}\end{equation}

Note that the term $\Phi _{{{\text{A}}_{qp,n,m}}}^{\text{LoS/NLoS}}\left( t \right)$ in (\ref{eq3})--(\ref{eq4}) is the time varying spatial phase related to movement direction and the UAV posture, which can be expressed as
\begin{equation}\begin{aligned}
  & \Phi _{{{\text{A}}_{qp,n,m}}}^{\text{LoS/NLoS}}\left( t \right)=k\left( \mathbf{r}_{p}^{\text{Tx}}\cdot {{\mathbf{R}}^{\text{Tx}}}\left( t \right)\cdot {{\mathbf{R}}^{\text{P}}}\left( t \right)\cdot \mathbf{s}_{\text{LoS/}n,m}^{\text{Tx}}\left( t \right) \right) \\ 
 & +k\left( \mathbf{r}_{q}^{\text{Rx}}\cdot {{\mathbf{R}}^{\text{Rx}}}\left( t \right)\cdot \mathbf{s}_{\text{LoS/}n,m}^{\text{Rx}}\left( t \right) \right)  
\end{aligned}
\label{eq8}\end{equation}
where $\mathbf{r}_{p}^{\text{Tx}}$ and $\mathbf{r}_{q}^{\text{Rx}}$ are the position vectors of the $p$-th transmitting antenna and the $q$-th receiving antenna, respectively, and the rotation matrix ${{\mathbf{R}}^{i}}\left( t \right)$, $i\in \left\{ \text{Tx},\text{Rx} \right\}$ aims to modify the position vector of Tx or Rx in real time \cite{ZhuQM20Sensors}, and can be expressed as
\begin{equation}
\begin{aligned}
  & {{\mathbf{R}}^{i}}\left( t \right) \\ 
 & \!\!=\!\!\left[ \begin{matrix}\!
   \cos {{\theta }^{i}}\left( t \right)\cos {{\phi }^{i}}\left( t \right) \!\!& \!-\!\sin {{\phi }^{i}}\left( t \right) \!\!&\!\! \!-\!\sin {{\theta }^{i}}\left( t \right)\cos {{\phi }^{i}}\left( t \right)  \\
   \cos {{\theta }^{i}}\left( t \right)\sin {{\phi }^{i}}\left( t \right)\!\! & \!\!\cos {{\phi }^{i}}\left( t \right)\!\! &\!\! \!-\!\sin {{\theta }^{i}}\left( t \right)\sin {{\phi }^{i}}\left( t \right)  \\
   \sin {{\theta }^{i}}\left( t \right) \!\!&\!\! 0 \!\!&\!\! \cos {{\theta }^{i}}\left( t \right)  \\
\end{matrix} \right] \\ 
\end{aligned}.
\label{eq9}
\end{equation}

To take the posture variation of UAV into account, a specific matrix ${{\mathbf{R}}^{\text{P}}}\left( t \right)$ is introduced in this paper. The time-variant roll angle $\omega \left( t \right)$, yaw angle $\varphi \left( t \right)$, and pitch angle $\gamma \left( t \right)$ are the rotated Euler angles when converting the world coordinate $x\text{-}y\text{-}z$ to the fuselage coordinate $\tilde{x}\text{-}\tilde{y}\text{-}\tilde{z}$ with respect to z axis, y axis, and x axis, respectively. In order to simplify the formulas, these time-variant parameters are marked as $\omega $, $\varphi $, and $\gamma $. Note that $\omega \in \left[ -\pi ,\pi  \right]$, $\varphi \in \left[ 0,2\pi  \right)$, and $\gamma \in \left[ -\pi ,\pi  \right]$. Then, the transfer matrices from the fuselage coordinate system to the world coordinate system are defined as
\begin{equation}{{\mathbf{R}}_{x}}=\left[ \begin{matrix}
   1 & 0 & 0  \\
   0 & \cos \left( \gamma  \right) & -\sin \left( \gamma  \right)  \\
   0 & \sin \left( \gamma  \right) & \cos \left( \gamma  \right)  \\
\end{matrix} \right]\label{eq10}\end{equation}
\begin{equation}{{\mathbf{R}}_{y}}=\left[ \begin{matrix}
   \cos \left( \varphi  \right) & 0 & \sin \left( \varphi  \right)  \\
   0 & 1 & 0  \\
   -\sin \left( \varphi  \right) & 0 & \cos \left( \varphi  \right)  \\
\end{matrix} \right]\label{eq11}\end{equation}
\begin{equation}{{\mathbf{R}}_{z}}=\left[ \begin{matrix}
   \cos \left( \omega  \right) & -\sin \left( \omega  \right) & 0  \\
   \sin \left( \omega  \right) & \cos \left( \omega  \right) & 0  \\
   0 & 0 & 1  \\
\end{matrix} \right].\label{eq12}\end{equation}
\newcounter{TempEqCnt} 
\setcounter{TempEqCnt}{\value{equation}} 
\setcounter{equation}{12} 
\begin{figure*}[htbp]
\normalsize
\begin{equation}\begin{aligned}
  & {{\mathbf{R}}^{\text{P}}}\left( t \right)={{\mathbf{R}}_{z}}{{\mathbf{R}}_{y}}{{\mathbf{R}}_{x}} \\ 
 & =\left[ \begin{matrix}
   \cos \left( \omega  \right)\cos \left( \varphi  \right) & \cos \left( \omega  \right)\sin \left( \varphi  \right)\sin \left( \gamma  \right)-\sin \left( \omega  \right)\cos \left( \gamma  \right) & \cos \left( \omega  \right)\sin \left( \varphi  \right)\cos \left( \gamma  \right)+\sin \left( \omega  \right)\sin \left( \gamma  \right)  \\
   \sin \left( \omega  \right)\cos \left( \varphi  \right) & \sin \left( \omega  \right)\sin \left( \varphi  \right)\sin \left( \gamma  \right)+\cos \left( \omega  \right)\cos \left( \gamma  \right) & \sin \left( \omega  \right)\sin \left( \varphi  \right)\cos \left( \gamma  \right)-\cos \left( \omega  \right)\sin \left( \gamma  \right)  \\
   -\sin \left( \varphi  \right) & \cos \left( \varphi  \right)\sin \left( \gamma  \right) & \cos \left( \varphi  \right)\cos \left( \gamma  \right)  \\
\end{matrix} \right] \\ 
\end{aligned}\label{eq13}\end{equation}
\end{figure*}
\setcounter{equation}{\value{TempEqCnt}}

Then the rotation matrix between the fuselage coordinate and the world coordinate can be calculated. The proposed term can be expressed as (\ref{eq13}), and the transformation can be obtained by
\setcounter{equation}{13} 
\begin{equation}
\left[ \begin{matrix}
   x  \\
   y  \\
   z  \\
\end{matrix} \right]={{\mathbf{R}}^{\text{P}}}(t)\left[ \begin{matrix}
   {\tilde{x}}  \\
   {\tilde{y}}  \\
   {\tilde{z}}  \\
\end{matrix} \right].
\label{eq14}
\end{equation}

Based on the posture matrix, the new model takes the posture variation of UAV into account, and thus can describe the arbitrary UAV posture rotations by setting the 3D rotational angles $\omega $, $\varphi $, and $\gamma $.

\section{STATISTICAL PROPERTY ANALYSIS OF PROPOSED MODEL}
\label{sec:Statistical Property Analysis of Proposed Model}
Due to the randomness of the channel, the statistical characteristics is generally used to evaluate its quality. When the channel model incorporates the fuselage posture, some typical channel statistical properties will change, thus their expressions need to be modified. First of all, the channel transfer function can be calculated by using Fourier transform on the corresponding time-variant CIR, i.e.,
\begin{equation}
\begin{aligned}
  & {{H}_{qp}}\left( \mathbf{r},f,t \right)=\int_{-\infty }^{\infty }{{{h}_{qp}}\left( t,\tau  \right){{e}^{-\text{j}2\pi f\tau }}}\text{d}\tau  \\ 
 & =\sqrt{\frac{K}{K+1}}{{B}^{\text{LoS}}}\left( t \right)h_{qp}^{\text{LoS}}\left( t \right){{e}^{-\text{j}2\pi f{{\tau }^{\text{LoS}}}\left( t \right)}} \\ 
 & +\sqrt{\frac{1}{K+1}}\sum\limits_{n=1}^{N\left( t \right)}{B_{n}^{\text{NLoS}}\left( t \right)h_{qp,n}^{\text{NLoS}}\left( t \right){{e}^{-\text{j}2\pi f\tau _{n}^{\text{NLoS}}\left( t \right)}}}  
\end{aligned}
\label{eq15}\end{equation}

, where $\mathbf{r}=\left\{ \Delta {{\mathbf{r}}^{\text{Tx}}},\Delta {{\mathbf{r}}^{\text{Rx}}} \right\}$ is the space lag, $\Delta {{\mathbf{r}}^{\text{Tx}}}=\mathbf{r}_{{{q}_{2}}}^{\text{Tx}}-\mathbf{r}_{{{q}_{1}}}^{\text{Tx}}$ and $\Delta {{\mathbf{r}}^{\text{Rx}}}=\mathbf{r}_{{{p}_{2}}}^{\text{Rx}}-\mathbf{r}_{{{p}_{1}}}^{\text{Rx}}$ denote the antenna element spacing of Tx and Rx, respectively. Then, by setting the frequency variation $\Delta f=0$, the spatial-temporal correlation function (ST-CF) of two specific CIRs ${{h}_{{{q}_{1}}{{p}_{1}}}}\left( t,\tau  \right)$ and ${{h}_{{{q}_{2}}{{p}_{2}}}}\left( t,\tau  \right)$ can be defined as
\begin{equation}\begin{aligned}
  & {{\rho }_{{{q}_{1}}{{p}_{1}},}}_{{{q}_{2}}{{p}_{2}}}\left( t,\mathbf{r};\Delta t,\Delta \mathbf{r} \right) \\ 
 & =E\left\{ H_{{{q}_{1}}{{p}_{1}}}^{*}\left( t,\mathbf{r} \right){{H}_{{{q}_{2}}{{p}_{2}}}}\left( t+\Delta t,\mathbf{r}+\Delta \mathbf{r} \right) \right\} \\ 
 & \text{=}\rho _{{{q}_{1}}{{p}_{1}},{{q}_{2}}{{p}_{2}}}^{\text{LoS}}\left( t;\Delta t,\left\{ \Delta {{\mathbf{r}}^{\text{Tx}}},\Delta {{\mathbf{r}}^{\text{Rx}}} \right\} \right) \\ 
 & \text{+}\rho _{{{q}_{1}}{{p}_{1}},{{q}_{2}}{{p}_{2}}}^{\text{NLoS}}\left( t;\Delta t,\left\{ \Delta {{\mathbf{r}}^{\text{Tx}}},\Delta {{\mathbf{r}}^{\text{Rx}}} \right\} \right)  
\end{aligned}\label{eq16}\end{equation}
where ${{\left( \cdot  \right)}^{*}}$ denotes the complex conjugation operation. Here, the local ST-CF can be calculated by the summation of LoS and NLoS components, which are assumed to be independent with each other. The detailed CF can be expressed as (\ref{eq17}) and (\ref{eq18}), where the effective phase term $\Phi _{qp}^{\text{LoS}}\left( t \right)$ and $\Phi _{qp,n,m}^{\text{NLoS}}\left( t \right)$ can be rewritten as
\begin{figure*}[htbp]
\normalsize
\begin{equation}
\rho _{{{q}_{1}}{{p}_{1}},{{q}_{2}}{{p}_{2}}}^{\text{LoS}}\left( t;\Delta t\left\{ \Delta {{\mathbf{r}}^{\text{Tx}}},\Delta {{\mathbf{r}}^{\text{Rx}}} \right\} \right)=\frac{K}{K+1}{{e}^{\text{j}\left( \Phi _{{{q}_{2}}{{p}_{2}}}^{\text{LoS}}\left( t+\Delta t \right)-\Phi _{{{q}_{1}}{{p}_{1}}}^{\text{LoS}}\left( t \right) \right)}}{{e}^{\text{j}2\pi f\left( {{\tau }^{\text{LoS}}}\left( t \right)-{{\tau }^{\text{LoS}}}\left( t+\Delta t \right) \right)}}\label{eq17}\end{equation}\end{figure*}
\begin{figure*}[htbp]
\footnotesize
\begin{equation}
\rho _{{{q}_{1}}{{p}_{1}},{{q}_{2}}{{p}_{2}}}^{\text{NLoS}}\left( t;\Delta t,\left\{ \Delta {{\mathbf{r}}^{\text{Tx}}},\Delta {{\mathbf{r}}^{\text{Rx}}} \right\} \right)=\frac{1}{M(K+1)}E\left\{ \sum\limits_{n=1}^{N(t)\cap N(t+\Delta t)}{\sum\limits_{m=1}^{M}{{{e}^{\text{j}\left(\! \Phi _{{{q}_{2}}{{p}_{2,n\!,\!m}}}^{\text{NLoS}}\left( t+\Delta t \right)-\Phi _{{{q}_{1}}{{p}_{1,n\!,\!m}}}^{\text{NLoS}}\left( t \right) \!\right)}}{{e}^{\text{j}2\pi f\left( \tau _{n}^{\text{NLoS}}\left( t \right)-\tau _{n}^{\text{NLoS}}\left( t+\Delta t \right) \right)}}}} \right\}
\label{eq18}\end{equation}
\end{figure*}
\begin{equation}\begin{aligned}
\Phi _{qp}^{\text{LoS}}\left( t \right)=\Phi _{\text{D}}^{\text{LoS}}\left( t \right)+\Phi _{{{\text{A}}_{qp}}}^{\text{LoS}}\left( t \right)
\end{aligned}\label{eq19}\end{equation}
\begin{equation}
\Phi _{qp,n,m}^{\text{NLoS}}\left( t \right)=\Phi _{{{\text{D}}_{n,m}}}^{\text{NLoS}}\left( t \right)+\Phi _{{{\text{A}}_{qp,n,m}}}^{\text{NLoS}}\left( t \right).
\label{eq20}\end{equation}

\subsection{Time-variant Auto Correlation Functions}
The temporal ACF is usually used to evaluate the time correlation of U2V channels and we can estimate the channel fading sensitivity with respect to the delay from the fluctuation of ACFs. The effect of survival probability from $t$ to $t+\Delta t$ should be taken into account and the ACF can be obtained by assuming $\Delta {{\mathbf{r}}^{\text{Tx}}}\text{=}0$ and $\Delta {{\mathbf{r}}^{\text{Rx}}}\text{=}0$, i.e., substituting ${{q}_{1}}\text{=}{{q}_{2}}\text{=}q$ and ${{p}_{1}}\text{=}{{p}_{2}}\text{=}p$ into (\ref{eq16}). The detailed expressions are shown as
\begin{equation}{{\rho }_{qp}}\left( t;\Delta t \right)\text{=}\rho _{qp}^{\text{LoS}}\left( t;\Delta t \right)\text{+}\rho _{qp}^{\text{NLoS}}\left( t;\Delta t \right)\label{eq21}\end{equation}
where
\begin{equation}
\begin{aligned}
  & \rho _{qp}^{\text{LoS}}\left( t;\Delta t \right) \\ 
 & =\frac{1}{K+1}{{e}^{\text{j}\left( \Phi _{qp}^{\text{LoS}}\left( t+\Delta t \right)-\Phi _{qp}^{\text{LoS}}\left( t \right) \right)}}{{e}^{\text{j}2\pi f\left( {{\tau }^{\text{LoS}}}\left( t \right)-{{\tau }^{\text{LoS}}}\left( t+\Delta t \right) \right)}} \\ 
\end{aligned}
\label{eq22}\end{equation}
and (\ref{eq23}), and $N(t)\cap N(t+\vartriangle t)$ denotes the set of shared paths at the time $t$ and $t+\Delta t$.
\begin{figure*}[htbp]
\small
\begin{equation}
\rho _{qp,n}^{\text{NLoS}}\left( \Delta t,t \right)=\frac{1}{M\left( K+1 \right)}E\left\{ \sum\limits_{n=1}^{N(t)\cap N(t+\Delta t)}{\sum\limits_{m=1}^{M}{{{e}^{\text{j}\left( \Phi _{q{{p}_{,n,m}}}^{\text{NLoS}}\left( t+\Delta t \right)-\Phi _{q{{p}_{,n,m}}}^{\text{NLoS}}\left( t \right) \right)}}{{e}^{\text{j}2\pi f\left( \tau _{n}^{\text{NLoS}}\left( t \right)-\tau _{n}^{\text{NLoS}}\left( t+\Delta t \right) \right)}}}} \right\}
\label{eq23}
\end{equation}
\end{figure*}

\subsection{Time-variant Cross-correlation Functions}
The spatial CCF can reflect the spatial correlation of the channel, especially the influence of change of antenna spacing on the U2V channel. By substituting $\Delta t\text{=}0$ into (\ref{eq16}), the normalized CCF between two different channel coefficients of proposed model can be obtained as
\begin{equation}
\begin{aligned}
  & {{\rho }_{{{q}_{1}}{{p}_{1}},}}_{{{q}_{2}}{{p}_{2}}}\left( \left\{ \Delta {{\mathbf{r}}^{\text{Tx}}},\Delta {{\mathbf{r}}^{\text{Rx}}} \right\};t \right) \\ 
 & \!\!=\!\rho _{{{q}_{1}}\!{{p}_{1}},{{q}_{2}}\!{{p}_{2}}}^{\text{LoS}}\!\left( \!\left\{ \Delta {{\mathbf{r}}^{\text{Tx}}}\!,\!\Delta {{\mathbf{r}}^{\text{Rx}}} \right\}\!;\!t \right)\!\text{+}\rho _{{{q}_{1}}\!{{p}_{1}},{{q}_{2}}\!{{p}_{2}}}^{\text{NLoS}}\!\left( \!\left\{ \Delta {{\mathbf{r}}^{\text{Tx}}}\!,\!\Delta {{\mathbf{r}}^{\text{Rx}}} \right\}\!;\!t \right)\! \\ 
\end{aligned}
\label{eq24}\end{equation}
where
\begin{equation}
\rho _{{{q}_{1}}{{p}_{1}},{{q}_{2}}{{p}_{2}}}^{\text{LoS}}\left( \left\{ \Delta {{\mathbf{r}}^{\text{Tx}}},\Delta {{\mathbf{r}}^{\text{Rx}}} \right\};t \right)\text{=}\frac{K}{K+1}{{e}^{\text{j}\left( \Phi _{{{q}_{2}}{{p}_{2}}}^{\text{LoS}}\left( t \right)-\Phi _{{{q}_{1}}{{p}_{1}}}^{\text{LoS}}\left( t \right) \right)}}\label{eq25}\end{equation}
\begin{equation}
\begin{aligned}
  & \rho _{{{q}_{1}}{{p}_{1}},{{q}_{2}}{{p}_{2}}}^{\text{NLoS}}\left( \left\{ \Delta {{\mathbf{r}}^{\text{Tx}}},\Delta {{\mathbf{r}}^{\text{Rx}}} \right\};t \right) \\ 
 & \text{=}\frac{1}{M\left( K+1 \right)}E\left\{ \sum\limits_{n=1}^{N(t)}{\sum\limits_{m=1}^{M}{{{e}^{\text{j}\left( \Phi _{{{q}_{2}}{{p}_{2,n,m}}}^{\text{NLoS}}\left( t \right)-\Phi _{{{q}_{1}}{{p}_{1,n,m}}}^{\text{NLoS}}\left( t \right) \right)}}}} \right\} \\ 
\end{aligned}
\label{eq26}\end{equation}

\section{NUMERICAL RESULTS AND DISCUSSIONS}
\label{sec:Numerical Results and Discussions}
\begin{figure}[H]
  \centering{\includegraphics[width=0.45\textwidth]{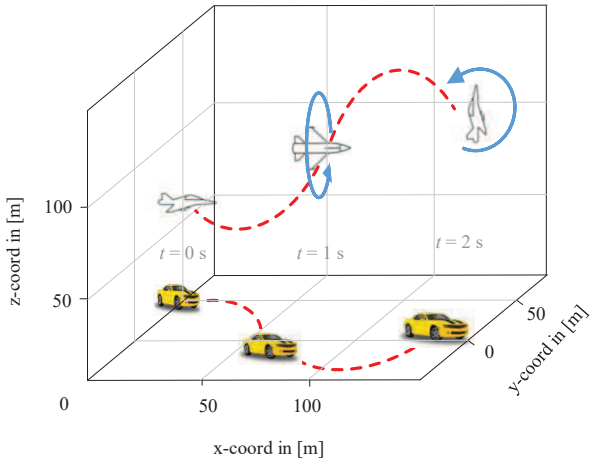}}
  \caption{the trajectories of UAV and vehicle.} 
  \label{fig2} 
\end{figure}

	In this section, some key channel statistical properties of proposed U2V channel model are studied. Especially, we investigate the impact of the rotation caused by the UAV posture rotations, and then verify the analytical statistical properties. The flight trajectory of UAV for simulations is shown in Fig. \ref{fig2}. It should be mentioned that the proposed model mainly focuses on the velocity and trajectory of ground terminal but does not consider the shape. Considering vehicles experience the most complicated movement on the ground, the other types of terminals, i.e., base station or pedestrian can be seen as special cases of vehicle. Therefore, we take the vehicle as an example in Fig. \ref{fig2}. The proposed model can be compatible with most air-to-ground communication scenarios through parameter modification.

Apparently, the movement of UAV not only includes the 3D arbitrary motion, but also the rotation movement. At the initial moment, the UAV is at an altitude of 150 meters with a speed of 50 m/s, without any rotation. Then, it rotates around an axis drawn through the body of the vehicle from tail to nose with an angular velocity ${\pi }/{2}\;\text{rad/s}$until the pitch angle reaches 90 degrees at $t$ = 1s. Also, from the moment of $t$ = 1s, the UAV begins to rotate upward related to the horizontal plane with an angular velocity ${\pi }/{2}\;\text{rad/s}$ until the roll angle reaches 90 degrees at $t$ = 2s. The vehicle on the ground has the absolute value of velocity as 20 m/s. Besides, the antenna pattern affects the small scale fading drastically. In the simulation, the radiation pattern of antenna array is assumed to obey 3D antenna model in the 3GPP TR standard \cite{3GPP2020} to make sure it is practical.

It should be noticed that we take the LoS path component and the first scattering component as examples to analyze the influence of time-variant fuselage posture on the channel characteristic, respectively. To highlight the impact of posture instead of the multipath propagation, the birth-death process of clusters in NLoS paths and the delay variations caused by moving of clusters are omitted. The simulated U2V communication system operates at 2.4 GHz and the path and sub-path angles are assumed to follow the 3GPP 38.901 standard definition \cite{ZhuQM21Wiley,3GPP2020}. The mean value of the Ricean K-factor is set as 7 and the variation is set as 4, which means the LoS component is also time-variant. 

\begin{figure}[htb]
  \centering{\includegraphics[width=0.45\textwidth]{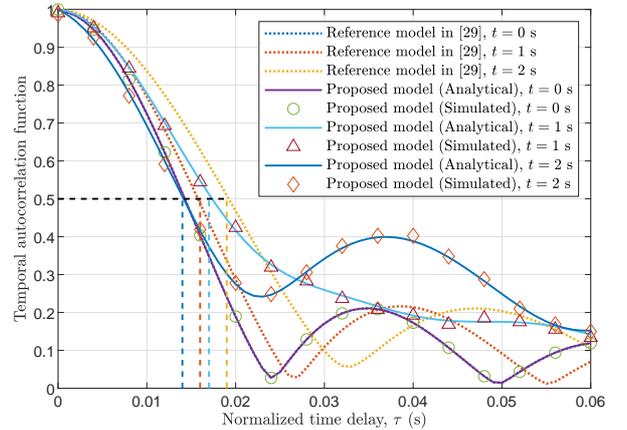}}
  \caption{ACFs of LoS component with/without the UAV posture.} 
  \label{fig3} 
\end{figure}

For the LoS case, the temporal ACFs of proposed model as well as the model in \cite{ZhuQM19MAP} which did not consider the UAV posture, are shown in Fig. \ref{fig3}. During the simulation, the moving direction of Tx and Rx and the fuselage posture change with time. From the figure, the values of ACFs are quite different at three moments, therefore the non-stationarity of U2V channel can be observed directly. The analytical results can be acquired by (\ref{eq22}), and the simulation results are obtained by calculating the correlation of channel CIRs. When $t$ = 0s, the ACFs of reference model and proposed model are precisely same, because there is no posture change of UAV. When $t$ = 1s and $t$ = 2s, the ACF of proposed model shows an obvious difference from the one of model in \cite{ZhuQM19MAP}. In addition, it can be seen that the simulated ACFs have a good consistency with the corresponding analytical results, which prove the correctness of derivations. 

Furthermore, the changing of UAV posture has certain impacts on the ACF. Different pitch and roll angles result in different trends of ACFs, which leads to the variation between the reference model and the proposed model. When $t$ = 1s, the pitch angle of UAV reaches 90 degrees. The posture matrix component ${{\mathbf{R}}_{x}}$ in (\ref{eq10}) has a predicable change as
\begin{equation}{{\mathbf{R}}_{x}}\left| _{\gamma =0} \right.=\left[ \begin{matrix}
   1 & 0 & 0  \\
   0 & 1 & 0  \\
   0 & 0 & 1  \\
\end{matrix} \right]\xrightarrow{\gamma \to \pi /2}\left[ \begin{matrix}
   1 & 0 & 0  \\
   0 & 0 & \text{-1}  \\
   0 & 1 & 0  \\
\end{matrix} \right]\label{eq27}\end{equation}
which leads to a time-variant rotation of posture matrix ${{\mathbf{R}}^{\text{P}}}$ and the effective phase term $\Phi _{qp}^{\text{LoS}}\left( t \right)$ in (\ref{eq19}), and eventually affects the channel ACFs. Likewise, when $t$ = 2s, the posture matrix component ${{\mathbf{R}}_{z}}$ changes as
\begin{equation}{{\mathbf{R}}_{z}}\left| _{\omega =0} \right.=\left[ \begin{matrix}
   1 & 0 & 0  \\
   0 & 1 & 0  \\
   0 & 0 & 1  \\
\end{matrix} \right]\xrightarrow{\omega \to \pi /2}\left[ \begin{matrix}
   0 & \text{-1} & 0  \\
   1 & 0 & 0  \\
   0 & 0 & 0  \\
\end{matrix} \right].\label{eq28}\end{equation}

Moreover, the coherence time, the minimum time lag when the ACF declines by half, can also be obtained directly from Fig. \ref{fig3}. With the increasing of pitch and roll angles, the coherence time becomes shorter. It can be found that the variation of pitch angles changes the coherence time from approximately 16 ms to 14 ms (when $t$ = 1s), while the variation of roll angles changes the coherence time from 19 ms to 17 ms (when $t$ = 2s). In terms of physical meaning, the additional posture rotation complicates the scattering environment which contributes to the decrease of coherence time.

\begin{figure}[htb]
  \centering{\includegraphics[width=0.45\textwidth]{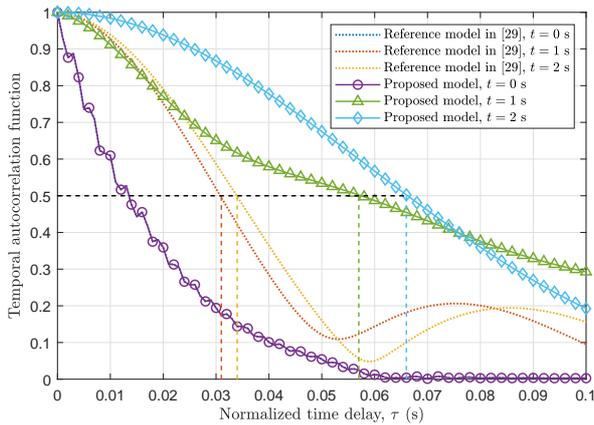}}
  \caption{ACFs of NLoS component with/without the UAV posture.} 
  \label{fig4} 
\end{figure}

Fig. \ref{fig4} shows the temporal ACFs of proposed model and the model in [29] in the NLoS case. It can be found that the variation of pitch angle leads the coherence time to change from approximately 31 ms to 57 ms (when $t$ = 1 s), while the variation of roll angle changes the coherence time from 34 ms to 66 ms (when $t$ = 2 s). Compared with the LoS case, the time-variant ACF is drastically affected and the coherence time becomes larger since the fuselage posture would aggravate the randomness of angles of departure and angles of arrival.

\begin{figure}[htb]
  \centering{\includegraphics[width=0.45\textwidth]{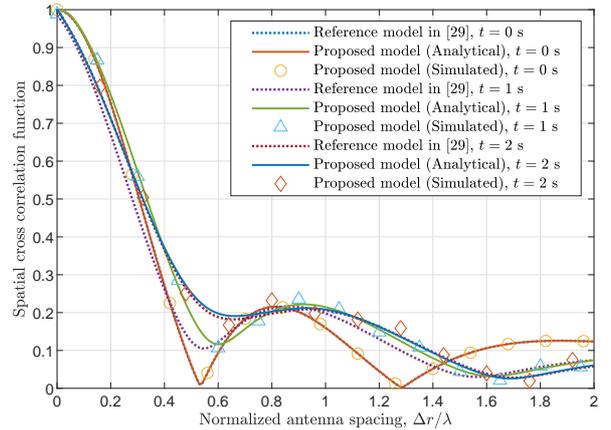}}
  \caption{CCFs of LoS component with/without the UAV posture.} 
  \label{fig5} 
\end{figure}

Under the same simulation scenario and trajectories, the spatial CCFs of proposed model as well as the reference model are shown in Fig. \ref{fig5}. We also give the analytical results which can be obtained by (\ref{eq25}) for comparison purpose. It can be observed that the simulated CCF provide a good match with the corresponding analytical results, which ensure the correctness of the derivations. Furthermore, the CCFs of reference model in \cite{ZhuQM19MAP} and that of the proposed model have a highly consistent trend, only with minor amplitude differences in some antenna spacing values. There is no evident variation between the reference model and the proposed model. It might be because the time-variant posture matrix ${{\mathbf{R}}^{\text{P}}}(t)$ is dominant temporal dependent, and varied slightly when the fuselage posture rotates. As a result, the effective phase term $\Phi _{qp}^{\text{LoS}}\left( t \right)$ and $\Phi _{qp,n,m}^{\text{NLoS}}\left( t \right)$ of different antenna pairs demonstrates a stability to antenna spacing. Thus, it can be inferred that the UAV posture rotation has slight impact on the values of CCFs, different pitch and roll angles of UAV will result in semblable trends of channel CCFs.

\begin{figure}[htb]
  \centering{\includegraphics[width=0.45\textwidth]{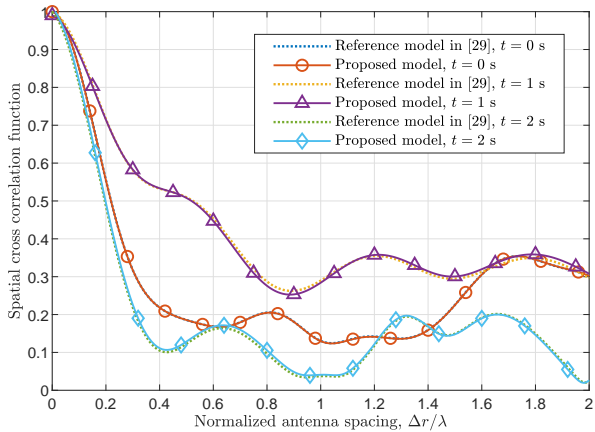}}
  \caption{CCFs of NLoS component with/without the UAV posture.} 
  \label{fig6} 
\end{figure}

As a supplementary argument to the inference above, the CCFs of proposed model and reference model under the NLoS scenario are shown in Fig. \ref{fig6}. It is obvious that the CCFs considering the fuselage posture have the same trends with the ones in \cite{ZhuQM19MAP}. The difference between two models is smaller than the one of LoS case. There are less varying values of LoS component compared to the NLoS one. The reason might be that the angles of arrival after scattering by the clusters are no longer mainly determined by the transmitting antenna on UAV in NLoS paths. The results reflect the relatively small angular spread and more stationary conditions in the LoS case. There are larger fluctuations in the NLoS case, which result in larger correlation variations. So in conclusion, the rotation of fuselage posture has merely a little effect on the CCFs.

\begin{figure}[htb]
  \centering{\includegraphics[width=0.45\textwidth]{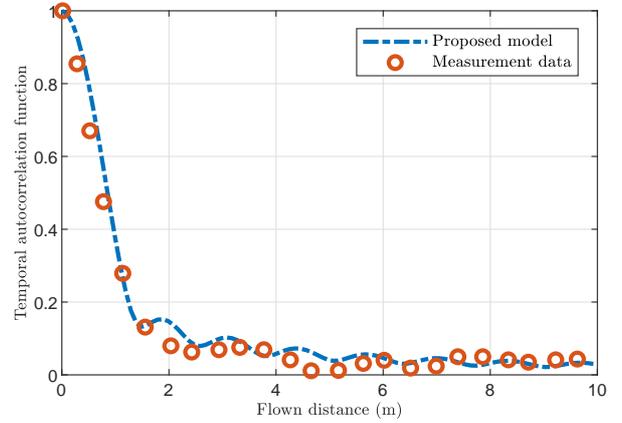}}
  \caption{ACF of proposed model and measurement data.} 
  \label{fig7} 
\end{figure}

\begin{figure}[ht]
  \centering{\includegraphics[width=0.45\textwidth]{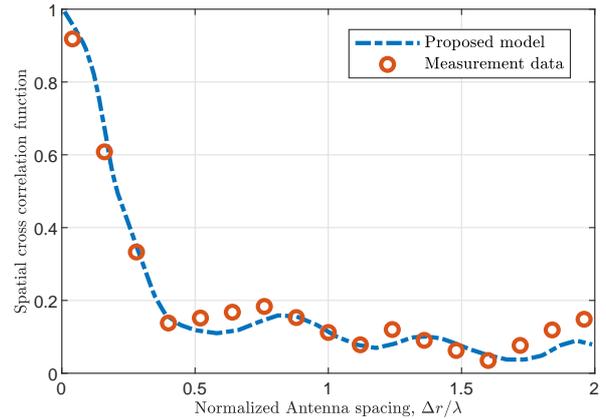}}
  \caption{CCF of proposed model and measurement data.} 
  \label{fig8} 
\end{figure}

To illustrate the compatibility of proposed model, we apply the channel parameters from a measurement campaign in \cite{Simunek13TAP}. The comparison of ACFs are shown in Fig. \ref{fig7}. The measured ACF is obtained as flown distances changed and the analytical value is obtained from the proposed model with the following parameters setting. The carrier frequency is 2.5 GHz and the speed of UAV and vehicle are 40 m/s and 10 m/s, respectively. The LoS path between the Tx and Rx is 1000 m at the beginning. The elevation angle of LoS path is ${\pi }/{3}\;$ and the azimuth angles of the speed vector are $\pi $ and ${\pi }/{4}\;$ respectively. It can be found that the ACF of proposed model matches well with the measurement data. Under the condition of the measurement campaign in \cite{Payami12EuCAP}, we match the analytical CCF with the measurement data as shown in Fig. \ref{fig8}. The analytical result is obtained with the following simulation parameters. The carrier frequency is 2.6 GHz and the LoS path is 500 m at the beginning. The elevation angle of LoS path is ${\pi }/{6}\;$and the azimuth angles of the velocity vectors $\pi $ and ${\pi }/{4}\;$ respectively. As shown in the figure, the proposed model also matches well with measurement results which testifies the generality of the proposed non-stationary U2V channel model. 

\section{CONCLUSIONS}
\label{sec:Conclusions}
In this paper, a realistic 3D non-stationary U2V channel model incorporating the fuselage posture has been proposed. The rotational movement caused by different fuselage postures has been considered by introducing the posture matrix. The analytical expressions of ACF and CCF have been derived and verified by simulation results. The analysis and simulation results have also shown that the UAV posture has significant impacts on the ACF. The CCF is less affected because the posture matrix is mainly temporal dependent. Moreover, the proposed GBSM can be applied to diverse UAV communication scenarios by adjusting model parameters, which is useful for the design, optimization, and evaluation of realistic UAV MIMO communication systems.

\section*{ACKNOWLEDGEMENT}
\label{ACKNOWLEDGEMENT}
This work was supported in part by the Fundamental Research Funds for the Central Universities (No.~NS2020026 and No.~NS2020063), in part by the Aeronautical Science Foundation of China (No.~201901052001), and in part by the National Key Scientific Instrument and Equipment Development Project under Grant (No.~61827801).

\end{document}